\documentclass[letter,traditabstract,longauth]{aa} 

\usepackage{graphicx}
\usepackage{txfonts}

\newcommand\T{\rule{0pt}{2.6ex}}
\newcommand\B{\rule[-1.2ex]{0pt}{0pt}}

\begin{document}
   \title{The {\it Herschel} Virgo Cluster Survey. III. A constraint on dust grain lifetime in early-type galaxies\thanks{ {\it Herschel} is an ESA space observatory with science instruments provided by European-led Principal Investigator consortia and with important participation from NASA.
}}
   \subtitle{}

\author{
M. S. Clemens\inst{1}
\and
A. P. Jones\inst{2}
\and
A. Bressan\inst{1,3}
\and
M. Baes\inst{4}
\and
G. J. Bendo\inst{5}
\and
S. Bianchi\inst{6}
\and
D. J. Bomans\inst{7}
\and
A. Boselli\inst{8}
\and
E. Corbelli\inst{6}
\and
L. Cortese\inst{9}
\and
A. Dariush\inst{9}
\and
J. I. Davies\inst{9}
\and
I. De Looze\inst{4}     
\and
S. di Serego Alighieri\inst{6}
\and
D. Fadda\inst{10}
\and
J. Fritz \inst{4}
\and
D. A. Garcia-Appadoo\inst{11}
\and
G. Gavazzi\inst{12}
\and
C. Giovanardi\inst{6}
\and
M. Grossi\inst{13}
\and
T. M. Hughes\inst{9}
\and
L. K. Hunt\inst{6}
\and
S. Madden\inst{14}
\and
D. Pierini\inst{15}
\and
M. Pohlen\inst{9}
\and
S. Sabatini\inst{16}
\and
M. W. L. Smith\inst{9}
\and
J. Verstappen\inst{4}
\and
C. Vlahakis\inst{17}
\and
E. M. Xilouris\inst{18}
\and
S. Zibetti\inst{19}
}

\institute{
INAF-Osservatorio Astronomico di Padova, Vicolo dell'Osservatorio 5, 35122 Padova, Italy\\
\email{marcel.clemens@oapd.inaf.it}
\and
Institut d'Astrophysique Spatiale (IAS), Batiment 121, Universite Paris-Sud 11 and CNRS, F-91405 Orsay, France\\
\email{Anthony.Jones@ias.u-psud.fr} 
\and
SISSA-ISAS, International School for Advanced Studies, Via Beirut 4, 34014 Trieste, Italy
\and
Sterrenkundig Observatorium, Universiteit Gent, Krijgslaan 281 S9, B-9000 Gent, Belgium 
\and
Astrophysics Group, Imperial College London, Blackett Laboratory, Prince Consort Road, London SW7 2AZ, UK 
\and
INAF-Osservatorio Astrofisico di Arcetri, Largo Enrico Fermi 5, 50125 Firenze, Italy 
\and
Astronomical Institute, Ruhr-University Bochum, Universitaetsstr. 150, 44780 Bochum, Germany 
\and
Laboratoire d'Astrophysique de Marseille, UMR 6110 CNRS, 38 rue F. Joliot-Curie, F-13388 Marseille, France 
\and
Department of Physics and Astronomy, Cardiff University, The Parade, Cardiff, CF24 3AA, UK
\and
NASA Herschel Science Center, California Institute of Technology, MS 100-22, Pasadena, CA 91125, USA 
\and
ESO, Alonso de Cordova 3107, Vitacura, Santiago, Chile 
\and
Universit\`{a} di Milano-Bicocca, piazza della Scienza 3, 20100, Milano, Italy 
\and
CAAUL, Observat\'orio Astron\'omico de Lisboa, Universidade de Lisboa,
Tapada da Ajuda, 1349-018, Lisboa, Portugal
\and
Laboratoire AIM, CEA/DSM- CNRS - Universit\'e Paris Diderot, Irfu/Service d'Astrophysique, 91191 Gif sur Yvette, France 
\and
Max-Planck-Institut fuer extraterrestrische Physik, Giessenbachstrasse, Postfach 1312, D-85741, Garching, Germany
\and
INAF-Istituto di Astrofisica Spaziale e Fisica Cosmica, via Fosso del Cavaliere 100, I-00133, Roma, Italy 
\and
Leiden Observatory, Leiden University, P.O. Box 9513, NL-2300 RA Leiden, The Netherlands 
\and
National Observatory of Athens, I. Metaxa and Vas. Pavlou, P. Penteli, GR-15236 Athens, Greece 
\and
Max-Planck-Institut fuer Astronomie, Koenigstuhl 17, D-69117 Heidelberg,  Germany 
}

   \date{Received ???; accepted ???}

 
  \abstract
    {Passive early-type galaxies (ETGs) provide an ideal laboratory for
studying the interplay between dust formation around evolved stars
and its subsequent destruction in a hot gas.
Using {\it Spitzer}-IRS and {\it Herschel} data we compare the dust production rate
in the envelopes of evolved AGB stars with a constraint on the total dust 
mass. Early-type galaxies which appear to be truly passively evolving 
are not detected by {\it Herschel}. We thus derive a distance independent upper 
limit to the dust grain survival time in the hostile environment of ETGs of 
$< 46\pm 25\;\rm Myr$ for amorphous silicate grains. This implies
that ETGs which \emph{are} detected at far-infrared wavelengths have acquired 
a cool dusty medium via interaction. Given likely time-scales for 
ram-pressure stripping, this also implies that only galaxies with dust
in a cool (atomic) medium can release dust into the intra-cluster medium.
}


   \maketitle
%

\section{Introduction}

In this paper we compare the dust outflow rate from the evolved AGB stars in passive ETGs, 
with the total dust mass, to estimate dust grain lifetimes. Both quantities are derived 
from observations; the outflow rate from {\it Spitzer}-IRS spectra, and the total dust mass
from {\it Herschel} (Pilbratt et al. 2010) maps of the {\it Herschel} Virgo Cluster Survey, HeViCS 
(Davies et al. 2010, www.hevics.org). We assume that evolved stars are the only source of 
dust in these passive systems as there is very little evidence that type Ia supernovae produce 
dust.

IRAS data have been used in the past in a similar way. Soifer et al. (1986) estimated the 
mass-loss from evolved stars in the bulge of M~31 from 12 and $25\;\rm \mu m$ fluxes and
the total dust mass from the 60 and $100\;\rm \mu m$ fluxes. They concluded that given 
that the observed dust mass could be produced by stellar outflows in
only $10^7\;\rm yr$, there must be some mechanism that depletes the bulge of inter-stellar 
matter. Jura et al. (1987) made similar calculations for a sample of elliptical galaxies. 
They estimated dust outflow rates by assuming that half of the $12\;\rm \mu m$ 
flux came from dusty stellar envelopes. They also found that the stellar mass-loss rate 
would be sufficient to produce the observed cool ISM in much less than a Hubble time. 

Dust mass production rate and total dust mass can only reasonably be
compared in objects that are truly passively evolving. Temi et
al. (2007) find, that for a given blue luminosity, the 70 and
$160\;\rm \mu m$ luminosities vary by 2 orders of magnitude,
indicating that many ETGs have a significant dust component not
directly attributable to mass-loss from evolved stars.  For our
purposes, these ETGs are not `passive' because other processes
(mergers?)  have probably contributed to the dust mass.

Recent studies of ETGs with {\it Spitzer} (Bressan et al. 2006;
Clemens et al. 2009) have shown that even in samples of ETGs selected
to be the most passive objects (ie. lying on the colour-magnitude
relation) a significant fraction show evidence of either on-going or
recent past star formation. These results indicate that ETGs in which
dust features are seen in the optical (e.g. Sadler \& Gerhard, 1985)
and many that have been detected by IRAS at 60 and $100\;\rm \mu m$
(Knapp et al. 1989) and {\it Spitzer} at 70 and $160\;\rm \mu m$
(Kaneda et al. 2007) may also host low levels of star
formation. Intriguigly, of the 7 ellipticals detected at 70 and
$160\;\rm \mu m$ by Kaneda et al. (2007) the only object not showing
PAH emission features is actually a radio source, so that the
far-infrared emission is probably synchrotron!
 
Tsai and Mathews (1995) studied dust destruction in ETGs via thermal sputtering in the
hot ($10^6 - 10^7$\,K),  low density ($n_H \sim 0.1$\,cm$^{-3}$) gas,  and found that
the destruction time-scale is short compared to any cooling  flow or dust transport
time-scale. They therefore concluded that dust is destroyed `on the spot' before it has
time to migrate within a galaxy. In a following work, Tsai and Mathews (1996)
found dust-to-gas mass ratios that are orders of magnitude less than that in typical stellar
ejecta or in the ISM of the Milky Way. Their interpretation of the IRAS data was, at that time,
hampered by the resolution of the existing observations; they were therefore not able to tie
down the origin and distribution of the observed dust emission.
 
As part of the {\it Herschel} Science Demonstration Phase (SDP), a $4\times4\;\rm deg^2$ field, 
centred approximately on M~87, has been observed as part of the HeViCS with both the PACS
(Poglitsch et al. 2010) and SPIRE (Griffin et al. 2010) instruments. See Davies et al. (2010)
 for details.

\section{Results}

Bressan et al. (2006) observed a sample of ETGs in the Virgo cluster 
with {\it Spitzer}-IRS. The objects were selected to lie on the optical colour-magnitude relation
and were expected to be passively evolving objects. The mid-infrared spectra for
this sample showed that only 76\% were actually passive objects, with the remainder showing
signs of either on-going or recent past star formation (PAH features) or AGN activity 
(atomic lines). Of this sample of 
17 galaxies, 9 lie in the HeViCS SDP field. 6 of these show IRS spectra consistent with 
passively evolving stellar populations (with the possible exception of NGC~4371), and it 
is these 6 objects, listed in Table~\ref{table:1}, 
on which we base our study. 

\subsection{Dust mass detection limit}

Of the 6 passive ETGs in the HeViCS SDP field, none are detected at any wavelength
 (100, 160, 250, 350 and $500\;\rm \mu m$)\footnote{We note that several ETGs \emph{are} 
detected in the SDP field, including M~87, M~84, M~86, NGC~4459, NGC~4435 
and NGC~4550, but none of these are truly passively evolving.}. 
Because of the sensitivity of the emission at the shorter PACS
wavelengths to small masses of relatively warm dust, we consider only the SPIRE images at 250, 
350 and $500\;\rm \mu m$ to determine the dust mass detection limit.  

Our mass detection limit depends on whether we consider the detection of an extended or 
point source. For passive galaxies, the dust should follow the stellar distribution, and 
thus be strongly centrally peaked.
As the core radius of an elliptical galaxy at the distance of the Virgo cluster
is typically less than the resolution of the SPIRE images (respectively 
$18^{\prime\prime}$, $25^{\prime\prime}$ and $37^{\prime\prime}$) the most likely detection 
would take the form of a point source.

Unless the dust is very cold ($\lesssim 17\;\rm K$) the tightest limit on the dust emission 
comes from the $250\;\rm \mu m$ map. This has a point source detection limit of 
$S_{\rm rms} = 9.6\;\rm mJy$. We calculate the dust mass detection limit as (Hildebrand, 1983):

\begin{equation}
M_{\rm d} \leq \frac{S_{\rm rms}(\nu) D^2}{\kappa_{\rm d}(\nu)B(\nu, T_{\rm d})}
\end{equation}

\noindent
where $D$ is the distance to the Virgo cluster (16.5 Mpc, Mei et al. 2007) and 
$\kappa_{\rm d}(\nu)$ is the dust mass opacity coefficient. The value of 
$\kappa_{\rm d}(\nu)$ depends on the grain composition but the mid-infrared spectra
(\S~\ref{dpragb}) are consistent with pure silicate grains. For dust 
composed only of amorphous silicate grains 
$\kappa_{\rm d}(250\;\rm \mu m) = 0.6517\;\rm m^2\,kg^{-1}$ 
(Draine \& Lee, 1984). Adopting this value, and $T_d = 30\,\rm K$, we arrive at a $2\sigma$ 
mass detection limit of $M_{\rm d} \leq 8.7 \times 10^3\;\rm M_{\odot}$\footnote{Were the
dust composed of amorphous carbon (Serra Diaz-Ca\~no \& Jones, 2008) then 
$\kappa_{\rm d}(250\;\rm \mu m) = 1.932$ (Rouleau \& Martin, 1991) and the mass 
detection limit $M_{\rm d} \leq 2.9 \times 10^3\;\rm M_{\odot}$.}.

\subsection{Dust production rate from AGB stars}
\label{dpragb}

Assuming a spherically symmetric, stationary stellar wind of velocity $v_\infty$, and 
dust mass-loss rate $\dot{M}_d^{CE}$ the circumstellar dust density is, 
\begin{equation}
\rho_d^{CE} = \frac{\dot{M}_d^{CE}}{4\pi r^2 v_\infty}
\label{eq1}
\end{equation}

\begin{figure}
\hspace{-10mm}
\includegraphics[width=6cm, angle=90]{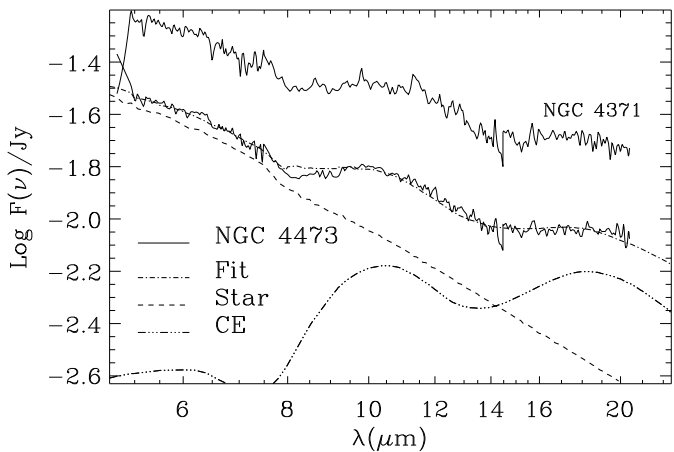}
   \caption{{\it Spitzer}-IRS spectrum of NGC~4473. The fitted model components are 
the stellar photosphere, and dusty circumstellar envelope (CE). The spectrum
of NGC~4371 is shifted by 0.2 dex.}
      \label{fig:1}
\end{figure}

The optical depth of the circumstellar envelope, $\tau$, is obtained by
multiplying Eq.~(\ref{eq1}) by the dust opacity coefficient and
integrating  from $r_{in}$ to $r_{out}$, where the former is 
the dust sublimation radius and the latter is a convenient outer radius\,$>>r_{in}$.
We thus have, 
\begin{equation}
\tau_1 =\frac{\dot{M}_d^{CE} k_1}{4\pi v_\infty r_{in} }
\label{tau}
\end{equation}
where $\kappa_1$ is the dust opacity at 1$\mu$m, following the notation of
Bressan et al. (1998), and we have neglected the very small term $1/r_{out}$.
The dust sublimation radius is $r_{in}\propto L^{1/2}$ (Granato \& Danese 1994) 
and the proportionality constant depends mainly on the dust composition
(through its sublimation temperature), the size distribution and only weakly
on the shape of the stellar spectrum. 
Inserting the value of $r_{in}$into Eq.~(\ref{tau}) we obtain,
\begin{equation}
\dot{M}_d^{CE}= A  \ \tau_1 \ v_\infty \ \sqrt{L_4} 
\label{MD}
\end{equation}
where $L_{4}$ is the bolometric luminosity in units of $10^{4}L_\odot$. 
The factor A is mainly determined by the dust mixture. For the mixture of silicate grains 
of M-type AGB stars used in Bressan et al. (1998), suitable for the spectra of passive ETGs 
(Fig.~\ref{fig:1}), A$\simeq\,6.5\times\,10^{-10}\; \rm M_{\odot}\,yr^{-1}/(km\,s^{-1})$.

To determine the dust mass-loss rate from the MIR spectrum we fit a combination of
a pure photospheric atmosphere, a MARCS model (Gustafsson et al. 2008) 
of 4000 K, plus the emission from a dusty envelope
selected from the library of Bressan et al. (1998), as described in Bressan et al. (2007).
We thus obtain the fraction of flux, say at 10$\mu$m,
$f_d^{\rm slit}$(10$\mu$m), due to dusty circumstellar envelopes, and, 
from the distance of the galaxy, the corresponding ``dust'' luminosity
sampled by the slit, $4\,\pi D^{2}f_d^{\rm slit}$(10$\mu$m).

Surprisingly, though ETGs contain a mixture of evolved stars
of different metallicity and age, a single dusty envelope
is typically enough for a good fit to the MIR spectrum.

The shape of the broad feature near 10$\mu$m in general requires
$\tau_1\sim\,$2--4, and so the envelopes are optically thin at 10$\mu$m.
From the {\it shape} of the 10$\mu$m feature we thus derive
the optical depth of the ``average'' AGB star. To compute its 
dust mass-loss rate through Eq.~(\ref{MD}) we need
to assume a bolometric luminosity and wind velocity.
Typical values for evolved, old stars are $L_4\sim$0.2--0.4 
and $v_\infty\sim 5-15\;\rm km\,s^{-1}$.

The total dust mass-loss sampled by the slit is finally obtained by multiplying the 
value for the ``average'' AGB star by the number of dusty AGB stars sampled by the 
slit. This is given by the ratio of the dust luminosity of the galaxy to that of the 
``average'' AGB star, $L_d^{CE}$(10$\mu$m). Thus, the total dust mass-loss sampled by 
the slit is,
 
\begin{equation}
\dot{M}_d^{\rm slit} = A  \ \tau_1 \ v_\infty \ \sqrt{L_4} \ \frac{4\,\pi D^{2}f_d^{\rm slit}(10\,\rm \mu m)}{L_d^{CE}(10\,\rm \mu m)}
\label{MDSLIT}
\end{equation}

Summarizing, the spectral fit provides $\tau_1$ and $f_d^{\rm slit}$(10$\mu$m).
From $\tau_1$, which characterizes the dusty envelope,
and the assumed bolometric stellar luminosity, we obtain $L_d^{CE}$(10$\mu$m) and finally
$\dot{M}_d^{\rm slit}$, for a specified value of $v_\infty$.
We note that, since  $L_d^{CE}$(10$\mu$m) scales approximately with the 
bolometric luminosity of the star, the total mass-loss has only a 
weak dependence on the assumed luminosity of the typical AGB star.
  
We finally note that recent work (Boyer et al. 2010) has shown dust to be released into 
the ISM only by AGB stars, contrary to some claims that RGB stars also contribute.

\begin{table}
\caption{Dust mass loss rates and grain lifetimes, $t_{\rm g}$.}
\label{table:1}
{\centering                    
\begin{tabular}{c c c c c c}       
\hline\hline                
NGC \T & $\dot{M}_{\rm d}/L_{\nu}(\rm K)^\dagger$ &$\dot{M}_{\rm d}/M^\ddagger$ & $\dot{M}_{\rm d}(\rm beam)$& $t_{g}^\ast$ \\
     \B & $\frac{\rm M_{\odot}yr^{-1}}{\rm 10^{26}\,W\,Hz^{-1}}$ &$\frac{\rm M_{\odot}yr^{-1}}{10^{12}M_{\odot}}$ &  $10^{-4}\;\rm M_{\odot}yr^{-1}$ & $10^{6}\;\rm yr$ \\
\hline
4371 \T & 3.80 & 0.00855 & 1.0  & 86\\
4442 & 3.48 & 0.00712 & 1.7  & 51\\
4473 & 3.57 & 0.00742 & 1.9  & 46\\
4474 & 3.71 & 0.00806 & 0.59 &150\\
4551 & 2.61 & 0.00468 & 0.36 &240\\
4564 & 4.28 & 0.00959 & 1.3  & 65\\
\hline                                  
\end{tabular}\\}
{\bf Notes.} $^\ast$ For an assumed dust temperature of 30~K. $^\dagger$ The specific dust 
mass-loss rate per unit K-band luminosity. $^\ddagger$ The mass loss per $10^{12}\;\rm M_{\odot}$ 
of galaxy mass. A dust mass opacity coefficient suitable for amorphous silicate grains has been 
used.
\end{table}

\subsection{Dust grain lifetimes}

Dust mass-loss rates were derived from {\it Spitzer}-IRS spectra that were taken through a slit
of dimensions $18^{\prime\prime} \times$ {3}\farcs{6} centred on the galaxy. In order to 
compare these values with the dust masses within one $250\;\rm \mu m$ beam we assume that 
the light distribution in the mid and far-infrared are similar, and that both are similar to 
that at K-band. For each galaxy, we use the 2MASS K-band image, smoothed to the {\it Spitzer}-IRS 
spatial resolution, to determine the flux within the IRS slit. We then smooth the K-band image 
to the {\it Herschel} $250\;\rm \mu m$ resolution and calculate the flux within one beam. We use the 
ratio of these two fluxes to derive the mass-loss rate from the area of one {\it Herschel} beam. 

We derive dust grain lifetimes simply as the ratio of dust mass detection limit to dust 
mass-loss rate, $M_{\rm d}/\dot{M}_d$. Our dust mass detection limit of 
$M_{\rm d} \leq 8.7 \times 10^3\;\rm M_{\odot}$ for silicate grains results in grain
lifetimes as given in Table~\ref{table:1}. The tightest constraint, of $<46\pm 25\;\rm Myr$, 
is provided by NGC~4473. The error reflects the uncertainties in $v_\infty$ and $L_4$. In 
Table~\ref{table:3} we provide the values used to arrive at 
this grain lifetime and their dependencies. For dust temperatures in the range $20-40\;\rm K$,
for example, the grain lifetime lies in the range $69-31\;\rm Myr$.

If dust grains are destroyed via sputtering in the $10^6 - 10^7\,\rm K$ gas in ETGs then the
theoretically, \emph{expected} dust grain lifetime in years is $t = k_i\, r_{\rm g}/n_{\rm H}$ 
(e.g., Tielens et al. 1994), where $r_{\rm g}$ 
is the grain radius in nm, $n_{\rm H}$ is the gas density in $\rm cm^{-3}$ and 
$k_{\rm i} = 310$ for silicate dust (210 for amorphous carbon, 1500 for graphite) so 
the grain lifetime should be longer in galaxies with a lower density hot ISM. 

Of our sample, NGC~4371 and NGC~4474 both lie in the least X-ray luminous category 
as defined by Irwin \& Sarazin (1998). NGC~4564 has been observed by Chandra, and the central 
hot gas density estimated to be $0.011\pm0.006\;\rm cm^{-3}$ by Soria et al. (2006). These 
authors analyzed a sample of `quiescent' galaxies and found central gas densities in the range 
0.011-$0.03\;\rm cm^{-3}$. These values overlap with the low end of estimates from Chandra 
data of more general ETG samples (Pellegrini, 2005). We thus assume a hot gas 
density of $0.02\;\rm cm^{-3}$ for all our ETGs. The expected dust grain lifetime, for
``typical'' $100\;\rm nm$ silicate grains is $1.55\;\rm Myr$ (1.05 for amorphous hydrocarbon, 
7.5 for graphite) and thus consistent with our non-detections.

\begin{table}
\caption{Parameter values for calculation of grain lifetime and the power of their 
dependence on the derived lifetime.}
\label{table:3}
{\centering                    
\begin{tabular}{c c c c c c c c}       
\hline\hline
$S_{\rm rms}$ \T  &$\kappa_{\rm d}(250)$&$T_{\rm d}$&$A^\ast$     &$\tau_1$&$v_{\infty}$& $L_4$\\ 
mJy              & $\rm m^2\,kg^{-1}$   & K         &$10^{-10}$ & &$\rm km\,s^{-1}$&$10^4L_{\odot}$\\
\hline

9.6  \T         &   0.6517            & 30           &$6.5$     & fit    & 10 & 0.3\\
 1               &   -1                & -1$^{\dagger}$ & 1        &  1     & 1 & 0.5 \\
\hline                                  
\end{tabular}\\}
{\bf Notes.} E.g., $t_{\rm g}\propto 1/T_{\rm d}$. $^\ast$ The units of $A$ are 
$\rm M_{\odot}\,yr^{-1}/(km\,s^{-1})$. $^{\dagger}$ In the Rayleigh-Jeans limit; 
which is a bad approximation for $T_{\rm d}\lesssim 20\;\rm K$.
\end{table}

\section{Discussion and conclusions}

We interpret the ratio of dust mass to dust production rate as a measure of grain
lifetime. In principle, however, dust could be removed from the galaxy rather than 
destroyed.

A wind of hot gas produced by type Ia supernovae probably cannot drive dust (or gas) out 
of the potential well of massive ETGs (Mathews \& Loewenstein 1986). However, good evidence
of ram-pressure stripping of the hot ISM in elliptical galaxies has been found in
the form of X-ray tails (Randall et al. 2008; Machacek et al. 2006). If dust is found in this 
hot gas then it too may be removed. However, the timescale for such stripping is estimated to 
be of the order a few $10^8\;\rm yr$ (Takeda et al. 1984; Murakami \& Babel 1999, eq. 12) which
is an order of magnitude longer than our upper limit to the grain lifetime. The observation
that the X-ray emission of most ETGs is \emph{not} displaced from the stellar distribution, 
in fact, suggestes that either the stripping time-scale is longer than the hot gas production 
time-scale or that the ram-pressure is typically insufficient to remove the hot medium. It is
now clear that if dust is isolated from the hot medium in denser, cooler clouds, than 
ram-pressure does have time to strip dust (Cortese et al. 2010) but the passive ETGs in our 
study do not show any evidence of accumulating dust in such clouds. \emph{Passive} ETGs do not 
directly pollute the intra-cluster medium with dust.

Grain lifetimes of $<46\;\rm Myr$ have immediate implications for ETGs that 
\emph{are} detected in the FIR. Either they produce dust at rates
much higher than those given in Table~\ref{table:1} or their dust is shielded from the hot gas in
cool clouds. For example, the dust mass of NGC~4435 of $1.2 \times 10^{6}\;\rm M_{\odot}$ would 
require a total dust production rate of $0.03\;\rm M_{\odot}\,yr^{-1}$, more than an order of 
magnitude greater than any value in Table~\ref{table:1}. This object is known to host star 
formation (Panuzzo et al., 2007) and so dust is likely to be in cool clouds more typical of 
late-type galaxies; grain lifetimes in this case may be much longer. Dust may have been 
acquired from an interaction, but in any case it probably arrived as part of a cooler 
medium.

In a passive ETG, the ISM, and the dust within it, come mainly from mass-loss from evolved 
stars. Unless the dust-to-gas ratio in this released material is very variable, one would 
expect dust and gas production to follow one another. If the X-ray luminosity, 
$L_{x}\propto n_{e}^2$ one would naively expect a relation
$L_{x}\propto M_{\rm d}^2$. However, if the grain lifetime, $t_{\rm g}\propto 1/n_{\rm H}$, then 
we would expect only a linear correlation between $L_{x}$ and $M_{\rm d}$. Kaneda et al. (2007) 
find evidence of an \emph{anti-correlation} for a small sample of elliptical galaxies 
detected by {\it Spitzer}-MIPS, but this may be expected if the dust were of external origin.

Only NGC~4371 is detected by MIPS at $70\;\rm \mu m$ (marginally at 
$160\;\rm \mu m$). The detection is consistent with our {\it Herschel} detection limits.
This object actually shows extremely weak PAH features in its {\it Spitzer}-IRS 
spectrum, and is therefore probably not totally passive. The object is included for comparison, and
its IRS spectrum included in Fig.~\ref{fig:1}.

We have assumed that AGB stars are the only source of dust. Were type Ia supernovae able to
produce dust from the metals they produce, we estimate that the contribution to the dust
production rate would be, at most, similar to that of AGB stars. Our limit on grain lifetime
would be a factor of $\sim$2 lower in this case.

Although dust grains are rather short-lived in ETGs, the prospects for survival in the 
intra-cluster  medium are rather better because the gas densities are typically at least 
2 orders of magnitude lower than in ETGs (B{\"o}hringer et al., 1994). Therefore, if dust 
can be removed from cluster galaxies, or form in the intra-cluster medium, there is a good 
chance that such dust will be detected by the HeViCS, especially at larger cluster radii 
where gas densities are lower.

\begin{acknowledgements}
MC and AB acknowledge support from contract ASI/INAF I/016/07/0.

\end{acknowledgements}

\end{document}